# 3D Meshes Registration : Application to statistical skull model

M. Berar [1], M. Desvignes[1], G. Bailly[2], Y. Payan[3]

[1] Laboratoire des Images et des Signaux (LIS), 961 rue de la Houille Blanche, BP 46, 38402 St. Martin d'Hères cedex, France
```
{Berar, Desvignes}@lis.inpg.fr
```
[2] Institut de la Communication Parlée (ICP), UMR CNRS 5009, INPG/U3, 46,av. Félix Viallet, 38031 Grenoble, France
```
Bailly@icp.inpg.fr
```
[3] Techniques de l'Imagerie, de la Modélisation et de la Cognition (TIMC), Faculté de Médecine, 38706 La Tronche, France
```
Payan@imag.fr
```

**Abstract.** In the context of computer assist surgical techniques, a new elastic registration method of 3D meshes is presented. In our applications, one mesh is a high density mesh (30000 vertexes), the second is a low density one (1000 vertexes). Registration is based upon the minimisation of a symmetric distance between both meshes, defined on the vertexes, in a multi resolution approach. Results on synthetic images are first presented. Then, thanks to this registration method, a statistical model of the skull is build from Computer Tomography exams collected for twelve patients.

## 1 Introduction

Medical Imaging and computer assisted surgical techniques may improve current maxillo-facial surgical protocol as an aid in diagnostic, planning and surgical procedure [1]. The steps of a complete assisted protocol may be summarized as : (1) Morphological data acquisition, including 3D imaging computed from Computer Tomography (CT) scanner, (2) Data integration which requires a 3D cephalometry analysis, (3) Surgical planning , (4) Surgical simulation for bone osteotomy and prediction of facial soft tissue deformation, (5) Per operative assistance for respecting surgical planning.
Three-dimensional cephalometric analysis, being essential for clinical use of computer aided techniques in maxillofacial, are currently in development [2,3,4].In most methods, the main drawback is the manual location of the points used to build the maxillofacial framework. The relationship between the cephalometry and the whole scans data is flawed by the amount of data and the variability of the exams. A common hypothesis is a virtual link between a low dimension model of the skull and these points.

We choose to first construct a statistical model of the skull, which will be link to a cephalometrics points model. This paper first presents data acquisition. In a second part, registration is described. Then, results on synthetic images are discussed and the construction of a statistical skull model is presented.

## 2 Method

The literature treating registration methods is very extensive (e.g., [5] for a survey). On one side are geometry based registration, which used a few selected points or features, where Iterative Closest Point and Active Shape Model are two classical approaches [6]. The main drawback of most of these methods is the need for the manual location of the landmarks used to drive the correspondence between objects in advance. On the other side are intensity-based algorithms, which use most of the intensity information in both data set [7].

### 2.1 Data Acquisition and 3D Reconstruction of the Patient's Skull

Coronal CT slices were collected for the partial skulls of 12 patients (helical scan with a 1-mm pitch and slices reconstructed every 0.31 mm or 0.48 mm). The Marching Cubes algorithm has been implemented to reconstruct the skull from CT slices on isosurfaces. The mandible and the skull are separated before the beginning of the matching process, our patients having different mandible relative position. (Figure 1, left panel).

In order to construct the statistical skull model, we need to register all the high density / low density meshes in a patient-shared reference system [8]. In this system, the triangles for a region of the skull are the same for all patients, the variability of the position of the vertexes will figurate the specificity of each density mesh in a patient. The vertex of these shared mesh can be considered as semilandmarks, i.e. as points that do not have names but that correspond across all the cases of a data set under a reasonable model of deformation from their common mean [9,10].

This shared mesh was not obtained with a decimation algorithm. Because our goal is to predict anatomical landmarks (some of cephalometric points) from the statistical skull model, we choose not to use a landmark based deformation [as in 11] but a method that does not require specification of corresponding features. The low definition model (Figure 1, right panel) was therefore taken from the Visible Woman Project.

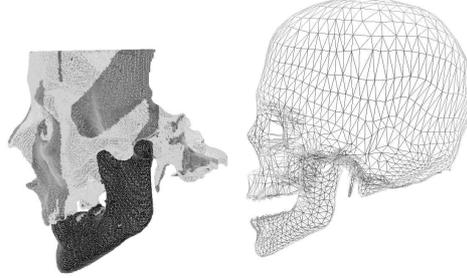

**Fig. 1. high definition mesh (*left*), low definition mesh(*right*).**

### 2.2 Shaping a generic model to patient-specific data : 3D Meshes registration

The deformation of a high definition 3D surface towards a low definition 3D surface is obtained by an original 3D-to-3D matching algorithm.

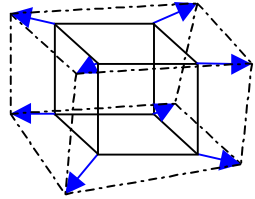

**Fig. 2. Applying a trilinear transformation to a cube**

#### 2.2.1 3D to 3D matching

The basic principle of the 3D-to-3D matching procedure developed by Lavallée and colleagues [12] consists basically in deforming the initial 3D space by a series of trilinear transformations applied to elementary cubes (see also figure 2 ) :

$$T_i(q,p) = \begin{bmatrix} p_{00} & p_{01} & & p_{07} \\ p_{10} & p_{11} & \cdots & p_{17} \\ p_{20} & p_{21} & & p_{27} \end{bmatrix} \begin{bmatrix} 1 & x_i & y_i & z_i & x_iy_i & y_iz_i & z_ix_i & x_iy_iz_i \end{bmatrix}^T \quad (1)$$

The elementary cubes are determined by iteratively subdividing the input space in a multi resolution scheme (see figure 3) in order to minimize the distance between the 3D surfaces:

$$\min_{p}\left[\sum_{i=1}^{N}\left[d\left(T(q_i,p),S\right)^2\right]+P(p)\right].\quad(2)$$

where S is the surface to be adjusted to the set of points q, p the parameters of the transformation T (initial rototranslation of the reference coordinates system and further a set of trilinear transformations). P(p) is a regularization function that guaranties the continuity of the transformations at the limits of each subdivision of the 3D space and that authorizes larger deformations for smaller subdivisions. The minimization is performed using the Levenberg-Marquardt algorithm [13].

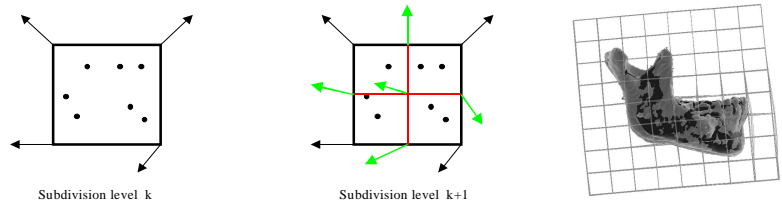

**Fig. 3. Subdivision of n elementary volume of the original space and new transformations vectors (2D simplification)** (*left*)**. Subdividing the space and applying the transformation** (*right*)**.**

**2.2.2 Symmetric distances**

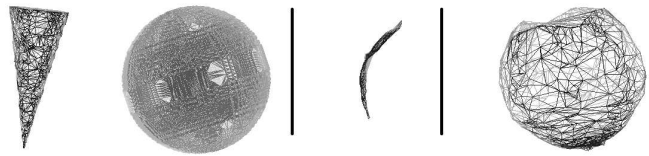

**Fig. 4. Matching a cone (source) toward a sphere (target)** (*left*)**. Mismatched cone using the single distance method** (*centre*)**; matched cone using the symmetric distance method** (*right*)**.**

In some cases, the transformed surface is well-matched to the closest surface but the correspondence between the two surfaces is false [see figure 4]. This mismatching can be explained by the two distances between each surfaces, which are not equivalent due to the difference of density between the two meshes. In this case, the distance from the source to the target (expressed in the minimization function) is very low whereas the distance from the target to the source is important (see Table 1).

We therefore included the two distances in the minimization function as in [14] :

$$\min_p \left[ \sum_{i=1}^{N} d(T(q_i,p), S_C)^2 + \sum_{i=1}^{N} d(T(q_i,p), bar(r_S))^2 + P(p) \right] \quad (3)$$

To compute the distance between the target and the source, the closest points of the low density vertexes towards the high density (points $q_i$ in equation 2) are stored. Bar($r_s$) is the barycentre of this set of points in the distance between the high density mesh (target) and the low density mesh (source).

**Table 1.** Evaluation of the two methods, matching a cone to a sphere

| Distances (mm) | Cone ->Sphere | | Sphere->Cone | |
|---|---|---|---|---|
| | mean | max. | mean | max. |
| Single | 0.15 | 1.55 | 18,03 | 36,42 |
| Symmetric | 0.29 | 3.79 | 0.72 | 7.81 |

## 3 Results

### 3.1 Synthetic images

We first try these two methods on a set of four forms obtained with the same procedure. Each form is generated with two levels of density (low and high) before or after decimation. The following table show the benefits of the "symmetric distance" method for these 8 objects, compared to the "single distance" method.

**Table 2 : Distance Gain (mm)**

| Target | Sphere | | Cube | | Open Ring | | Cone | |
|---|---|---|---|---|---|---|---|---|
| Source | Low | High | Low | High | Low | High | Low | High |
| Sphere low | | 0 | -0,17 | 9,77 | -0,1 | 4,38 | 4,9 | 4,99 |
| Sphere high | 0 | | 0,55 | -0,19 | -0,3 | 0,09 | 2,58 | 2,94 |
| Cube low | 2,1 | 3,58 | | 0,44 | 3,2 | 5,92 | 20,06 | 17,83 |
| Cube high | -1,3 | -0,5 | 0 | | 6,63 | 5,74 | 9,54 | 8,48 |
| Open Ring low | 24,16 | 21,75 | -0,05 | 3,72 | | 0 | 13,94 | 15,02 |
| Open Ring high | 13,02 | 16,26 | -0,01 | 0 | 0 | | 4,5 | 12,41 |
| Cone Low | 26,41 | 28,61 | 14,54 | 25,41 | 4,4 | 5,63 | | 0 |
| Cone high | 11,99 | 21,69 | 6,04 | 9,54 | 1,67 | 1,11 | -0,01 | |

Table 2 summarises results : The method is well suited for shapes of same topology. But different topologies are not registered: a sphere deformed to the open ring shape will not capture the aperture of the ring, and a cone will "flat" himself in the centre of the ring.

### 3.2 Real Data : Mandible Meshes

The low density mandible meshes are generated using the "symmetric distance" method. The single distance approach leads to many mismatches in the condyle and goniac angle regions (figure 5).

The maximal distances are located on the teeth (which will not be included in the model, but are used for correspondences during the registration) and in the coronoid regions.

The mean distances can be considered as the registration noise, due to the difference of density (see Table 3).

**Table 3** : Mean distances between meshes

| Distances (mm) | Low->High | | High->Low | |
|---|---|---|---|---|
| | mean | max. | mean | max. |
| Single | 1.27 | 9.28 | 5.80 | 56.87 |
| Symmetric | 1.33 | 8.42 | 2.57 | 22.78 |

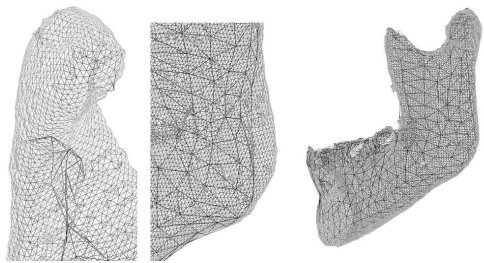

Fig. 5. **mismatched parts of mandible using the single distance method (*left* : condyle, *center* : goniac angle) and matched low density mesh to high density mesh using symmetric distance method.**

### 3.3 Application : Skull Statistical Model

12 CT patient's scans with different pathologies are used. Half of them suffer from sinus pathologies, while the other half suffer from pathology of the orbits. The CT scans are centred around the pathology and do not include (except for one patient) the skull vault. The patients have different mandible positions, so the skull and the mandible were registered separately.

After jointing these two parts of our model, they are aligned using Procrustes registration on the mean individual, as the statistical shape model must be independent from the rigid transformations (translation, rotation). Gravity centres are first aligned. Then the optimal rotation that minimizes the distance between the two set of points is obtained.

The statistical model can only have 12 degrees-of-freedom (DOF), for a set of 3938 points (potentially 11814 geometrical DOF), as the number of DOF is limited by the number of patients. Using a simple statistical analysis, we show that 95% of the variance of the data can be explained with only 5 parameters (see Table 4). These "shape" parameters are linear and additive :

$$P = M + A*\alpha. \quad (4)$$

where M is the mean shape, A the "shape" vector, and $\alpha$ the shape coefficients.

**Table 4 :** variance explained by parameters

| Parameter | 1 | 2 | 3 | 4 | 5 |
|---|---|---|---|---|---|
| Variances % | 52,11 | 19,81 | 11,14 | 9,55 | 2,97 |
| Cumulated Variance % | 52,11 | 71,92 | 83.06 | 92.61 | 95.58 |

Figure 6 shows the effects of the two first parameters. The first parameter is linked to a global size factor, whereas the second influences the shapes of the forehead and of the cranial vault.

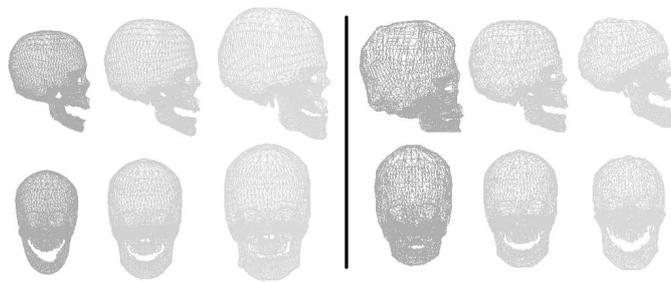

Fig. 6. **Effects of the first (*left*) and second (*right*) parameters for 3 times the standard deviations.**

## 4 Conclusion

In this paper, a new registration approach for 3D meshes has been presented. In our application, one mesh is a high density mesh, the second a low density one. To enhance the registration, a symmetric distance has been proposed in a multi resolution approach. Results on synthetic and real images exhibit good qualitative performances. This method is then used to elaborate a statistical skull model.